\documentclass[aps,prd,preprint,onecolumn,superscriptaddress,nofootinbib,floatfix,longbibliography]{revtex4-2}

\usepackage{amsmath}
\usepackage{placeins}
\usepackage{physics}
\usepackage{xcolor}
\usepackage{graphicx}
\usepackage{dcolumn}
\usepackage{bm}
\usepackage[caption=false]{subfig}

\usepackage{comment}

\begin{document}

\title{Influence of dark matter on the formation of biogenic elements in early universe stars}

\author{L. Yıldız}
\email{li.yildiz.na@gmail.com}
\author{D. Kaykı}
\email{dehakayki.science.technology@gmail.com}
\author{M. F. Ciappina}
\email{marcelo.ciappina@gtiit.edu.cn}
\affiliation{Department of Physics, Guangdong Technion - Israel Institute of Technology, 241 Daxue Road, Shantou, Guangdong, 515063, China}
\affiliation{Technion - Israel Institute of Technology, Haifa, 32000, Israel}
\affiliation{Guangdong Provincial Key Laboratory of Materials and Technologies for Energy Conversion, Guangdong Technion - Israel Institute of Technology, 241 Daxue Road, Shantou, Guangdong, 515063, China}

\begin{abstract}
We demonstrate that DM interactions can profoundly influence stellar nucleosynthesis in the early universe by altering thermodynamic gradients and modifying nuclear reaction rates within primordial stars. Incorporating a DM–modified Fermi-Dirac distribution and accounting for localized energy injection from annihilation heating, our model predicts enhanced production of carbon and nitrogen alongside reduced oxygen synthesis. These compositional shifts significantly reshape stellar structure and produce synthetic spectra that closely reproduce the observed characteristics of carbon-enhanced metal-poor (CEMP) stars. Our findings reveal a direct and previously overlooked role of DM in driving the chemical evolution of the early cosmos, offering a plausible link between fundamental particle physics and observable astrophysical signatures.
\end{abstract}

\maketitle
\clearpage
\section{Introduction}
The synthesis of biogenic elements—carbon (C), nitrogen (N), and oxygen (O)—in the first generation of stars is not merely a matter of stellar evolution; it is a fundamental problem at the interface of nuclear astrophysics, cosmology, and particle physics. The canonical theory of stellar nucleosynthesis, while successful in describing equilibrium fusion pathways in main-sequence and evolved stars, fails to reproduce the peculiar abundance patterns observed in ancient, metal-deficient stars known as carbon-enhanced metal-poor (CEMP) stars, particularly the CEMP-no subclass lacking neutron-capture enrichment \cite{Beers2005, Frebel2015}. 

These stars often exhibit elevated carbon-to-oxygen (C/O) and carbon-to-nitrogen (C/N) ratios that lie outside the range predicted by standard stellar evolution tracks \cite{Meynet2006, Yoon2016}. The persistence of these anomalies across independent surveys indicates a missing physical ingredient during the formation of Population III stars. Simultaneously, DM (DM) continues to elude direct detection despite its overwhelming gravitational signature on galactic and cosmological scales. Accounting for roughly $85\%$ of the matter content of the Universe, DM remains undetected in laboratory settings, leaving its interaction cross-section with baryonic matter effectively unconstrained \cite{Bertone2005, Khlopov2011}. 

Theoretical models predict that Weakly Interacting Massive Particles (WIMPs), the prototypical DM candidates, may be gravitationally captured by stars during formation, especially within dense minihalos in the early Universe \cite{Iocco2008, Spolyar2008}. Once thermalized, these particles could self-annihilate and deposit energy within the stellar core, thereby altering local thermodynamic conditions and triggering non-standard nuclear pathways .

In this work, we propose a fully coupled theoretical and numerical framework that integrates DM microphysics into the stellar structure equations of primordial stars. Unlike earlier models that treat DM effects as perturbative or external sources, our approach directly embeds the influence of DM annihilation, gravitational compression, and quantum statistical corrections into the core thermodynamics of Population III stars. This framework allows us to probe how DM can reshape stellar evolution from first principles and provides a viable mechanism for resolving the long-standing CEMP-no abundance anomalies. 

The remainder of this paper is structured as follows. In Section~\ref{sec:theory}, we formulate the full theoretical model incorporating DM annihilation, gravitational effects, and quantum-statistical corrections. Section~\ref{sec:methods} presents the numerical methods and implementation details. Section~\ref{sec:results} provides our simulation results, with a focus on DM-induced modifications to thermodynamic profiles, nucleosynthetic yields, and spectral signatures. Here, we also contextualize our findings within the broader landscape of stellar evolution, DM astrophysics, and observational data. Finally, Section~\ref{sec:conclusion} summarizes our conclusions and outlines potential avenues for future work.

\section{Theoretical Framework}
\label{sec:theory}

We develop a first-principles extension of classical stellar structure theory by explicitly incorporating gravitationally bound and thermally equilibrated DM into primordial stellar interiors \cite{Khlopov2011, DiCintio2014, Gnedin2004}. This extended model modifies the standard equations governing hydrostatic equilibrium and energy transport through a gravitational potential resulting from the combined distribution of baryonic and DM components. All derivations and analyses are strictly confined to stellar scales, deliberately excluding galactic-scale phenomena such as the core–cusp problem in DM halos. Our framework consistently integrates the DM density profile, annihilation-driven energy injection, gravitational coupling, and quantum-statistical corrections into the stellar structure equations, forming a coherent theoretical foundation to investigate Population III stars significantly influenced by DM.

The DM distribution within the stellar interior is modeled using a generalized Navarro–Frenk–White (gNFW) profile:
\begin{equation}
\rho_\chi(r) = \rho_0 \left( \frac{r}{r_s} \right)^{-\gamma} \left(1 + \frac{r}{r_s} \right)^{\gamma - \beta},
\end{equation}
where $\rho_0$ is the central DM density, $r_s$ is the scale radius, and $(\gamma, \beta)$ are the inner and outer slope parameters, respectively. We adopt $\gamma = 1.2$ and $\beta = 3$, consistent with simulations of baryon-compressed minihalos~\cite{DiCintio2014, Gnedin2004}. These values reflect the steepening of the inner DM density profile due to adiabatic contraction during primordial gas collapse. For Population III environments, we use fiducial values $\rho_0 = 10^{-23}~\mathrm{cm^{-3}}$
and $r_s = 10^9~\mathrm{cm}$, corresponding to expected conditions in minihalos at redshifts $z \sim 20$~\cite{Spolyar2008}. This formulation captures local, stellar-scale effects of DM and should not be conflated with large-scale halo profiles related to the core–cusp problem. The term "generalized NFW" refers to this adjustable-slope version of the standard NFW profile, which is recovered for $(\gamma, \beta) = (1, 3)$~\cite{Navarro1996}.

We assume DM consists of self-annihilating Majorana fermions, consistent with WIMP models. As these particles become gravitationally captured within the protostellar potential well and thermalize, they can annihilate and deposit energy into the stellar core. The local volumetric energy deposition rate is given by:
\begin{equation}
Q_\chi(r) = \frac{\langle \sigma v \rangle}{m_\chi} \rho_\chi^2(r),
\label{eq:heating}
\end{equation}
where $\langle \sigma v \rangle = 3 \times 10^{-26}~\mathrm{cm^3\,s^{-1}}$ is the thermal annihilation cross-section, and $m_\chi = 100~\mathrm{GeV}$ is the DM mass. The cumulative luminosity due to annihilation is computed via:
\begin{equation}
L_\chi(r) = \int_0^r 4\pi r'^2 Q_\chi(r')\,dr'.
\end{equation}
This heating source modifies the radiative energy transport equation \cite{Freese2009, Raffelt1996}, which becomes:
\begin{equation}
\frac{dT}{dr} = -\frac{3\kappa\rho}{16\pi a c T^3 r^2}(L_r - L_\chi(r)),
\label{eq:temperature-gradient}
\end{equation}
where $\kappa$ is the Rosseland mean opacity, $a$ is the radiation constant, and $c$ is the speed of light.

In addition to modifying energy transport, DM also affects the quantum statistical behavior of the baryonic plasma. We account for this via a spatially varying chemical potential shift in the Fermi–Dirac distribution:
\begin{equation}
f(E, r) = \frac{1}{\exp\left[ \frac{E - \mu_0 - \delta\mu_\chi(r)}{k_B T(r)} \right] + 1},
\end{equation}
where $\mu_0$ is the unperturbed chemical potential and $\delta\mu_\chi(r) = \alpha_\chi \Phi_\chi(r)$ represents the gravitationally induced shift. The potential $\Phi_\chi(r)$ is obtained by solving Poisson’s equation:
\begin{equation}
\nabla^2\Phi_\chi(r) = \frac{1}{r^2}\frac{d}{dr}\left(r^2 \frac{d\Phi_\chi}{dr}\right) = 4\pi G\rho_\chi(r),
\end{equation}
with boundary conditions ensuring regularity at the center and decay at the stellar surface. The dimensionless parameter $\alpha_\chi$ encodes the sensitivity of baryons to local gravitational variations at the level of occupation statistics, and is chosen to maintain thermodynamic stability within the degenerate regime. The net effect of $\delta\mu_\chi$ is to enhance degeneracy pressure and modify reaction thresholds—particularly the triple-alpha and $^{12}\mathrm{C}(\alpha,\gamma){}^{16}\mathrm{O}$ processes—in a manner sensitive to the steepness of the local gravitational potential.

To account for DM’s gravitational influence on hydrostatic equilibrium, we generalize the Lane–Emden equation under the assumption of a polytropic equation of state $P = K \rho^{1 + 1/n}$. We define dimensionless variables:
\begin{equation}
\xi = \frac{r}{a}, \quad \theta(\xi) = \left[\frac{\rho(r)}{\rho_c}\right]^{1/n}, \quad a = \sqrt{\frac{(n+1)K}{4\pi G}}\rho_c^{\frac{1-n}{2n}},
\end{equation}
and modify the classical equation to include a DM source term:
\begin{equation}
\frac{1}{\xi^2} \frac{d}{d\xi} \left( \xi^2 \frac{d\theta}{d\xi} \right) = -\theta^n(\xi) - \beta_\chi \psi(\xi),
\end{equation}
where $\psi(\xi)$ is the scaled gNFW profile and $\beta_\chi = \frac{4\pi G \rho_{0,\chi} a^2}{K(n+1)}$ quantifies the DM gravitational contribution. The system is solved numerically using a fourth-order Runge–Kutta–Fehlberg scheme with adaptive mesh refinement. Here, the boundary conditions are set to $\theta(0)=1$ and $d\theta/d\xi|_{\xi=0}=0$, and solutions are verified for dimensional consistency and convergence below $10^{-6}$ in temperature and density.

This formulation creates an effective two-fluid gravitational potential, modifying the inner potential well, especially near the core where $\rho_\chi(r)$ is steepest. The resulting structural deformation affects core temperatures, fusion thresholds, and chemical yields. Our framework, applied to stellar masses in the range $0.7 - 1.2\,M_\odot$, reveals enhanced triple-alpha burning, suppressed oxygen synthesis via modified Fermi–Dirac occupation, and chemical yields consistent with CEMP-no stars like HE0107–5240 and BD+44$^\circ$493.

It is crucial to emphasize that this model is confined to stellar-scale phenomena and does not aim to resolve the galactic-scale cusp–core problem. The cuspy nature of the DM profile arises from local gravitational compression and adiabatic contraction during star formation in minihalos. The simulated chemical patterns and thermodynamic responses support the view that Population III stars may serve as astrophysical laboratories for probing DM microphysics.

\section{Methods}
\label{sec:methods}

This section presents the numerical methods used to solve the modified stellar structure equations introduced in Sec.~\ref{sec:theory}. Our approach consists of discretizing the modified Lane–Emden equation, implementing a high-accuracy Runge–Kutta integrator, and verifying the physical consistency and convergence of the solution under varying DM parameters.

\subsection{Numerical Integration of the Modified Lane--Emden Equation}

The modified Lane–Emden equation derived in Eq.~\eqref{eq:modle-repeat} incorporates a source term $f(\xi)$ that encodes the effects of DM density. The equation takes the form:
\begin{equation}
\frac{1}{\xi^2} \frac{d}{d\xi} \left( \xi^2 \frac{d\theta}{d\xi} \right) = -\theta^n(\xi) - \beta f(\xi),
\label{eq:modle-repeat}
\end{equation}
with boundary conditions:
\begin{equation}
\theta(0) = 1, \quad \left. \frac{d\theta}{d\xi} \right|_{\xi=0} = 0.
\end{equation}
We solve this equation using a fourth-order Runge–Kutta method with adaptive step-size control and a fixed absolute error tolerance $\epsilon = 10^{-6}$. Near the singular point $\xi = 0$, we avoid numerical instability by using a second-order Taylor expansion:
\begin{equation}
\theta(\xi) = 1 - \frac{1}{6} \left(1 + \beta f(0) \right)\xi^2 + \mathcal{O}(\xi^4).
\end{equation}
This expansion provides the initial conditions for the initial integration steps.

\subsection{Mesh and Discretization Parameters}

We construct a radial mesh in the dimensionless coordinate $\xi \in [0, \xi_{\mathrm{max}}]$, where $\xi_{\mathrm{max}}$ corresponds to the radius at which $\theta(\xi)$ vanishes (i.e., the stellar surface). Unless otherwise specified, we use $N = 10^4$ grid points with uniform spacing $\Delta \xi = \xi_{\mathrm{max}}/N$. Mesh convergence was tested with $N = 10^3$, $10^4$, and $10^5$ to ensure independence from numerical resolution.

The DM density source term $f(\xi)$ is precomputed using the gNFW profile as:
\begin{equation}
f(\xi) = \left( \frac{\xi}{\xi_s} \right)^{-\gamma} \left(1 + \frac{\xi}{\xi_s} \right)^{\gamma - \beta},
\end{equation}
with $\xi_s = r_s / a$. This function is normalized such that $f(0) = 1$ and tabulated for numerical interpolation.

\subsection{DM Coupling Strength and Parameters}

The coupling constant $\beta$ in Eq.~\eqref{eq:modle-repeat} governs the strength of gravitational interaction between baryons and the localized DM distribution. It is defined as:
\begin{equation}
\beta = \frac{4\pi G a^2 \rho_{0,\chi}}{P_c},
\end{equation}
where $G$ is the gravitational constant, $\rho_{0,\chi}$ is the central DM density from the gNFW profile, $a$ is the Lane–Emden radial scaling factor, and $P_c$ is the central pressure of the baryonic gas.
Furthermore, the scaling factor $a$ is given by:
\begin{equation}
a = \sqrt{\frac{(n+1)K}{4\pi G}}\, \rho_c^{\frac{1-n}{2n}},
\end{equation}
where $K$ is the polytropic constant, $\rho_c$ is the central baryonic density, and $n$ is the polytropic index. Likewise, the central pressure $P_c$ is defined as:
\begin{equation}
P_c = K \rho_c^{1 + 1/n}.
\end{equation}
These expressions ensure that $\beta$ remains a dimensionless parameter quantifying the relative strength of DM gravity in the context of stellar hydrostatics. We vary $\beta$ in the range $[0, 0.3]$ to explore different levels of DM concentration and its thermodynamic consequences.

\subsection{Code Validation and Consistency Checks}

To validate our implementation, we first consider the limit $\beta = 0$ in which the modified Lane–Emden equation reduces to its classical form. In this case, the numerical solutions for $n = 1.5$ and $n = 3$ exactly reproduce known analytical and tabulated results for polytropic stellar models. Convergence testing is performed by monitoring variations in $\theta(\xi)$, the stellar radius $\xi_R$, and derived physical quantities such as central temperature $T_c$ and density $\rho_c$ under successive mesh refinements. We confirm that all quantities converge within a fractional error $\delta < 10^{-5}$.

\subsection{Energy Injection and Thermodynamic Feedback}

Using the numerical solution $\theta(\xi)$, we reconstruct the radial density and temperature profiles and compute the annihilation heating rate:
\begin{equation}
Q_\chi(r) = \frac{\langle \sigma v \rangle}{m_\chi} \rho_\chi^2(r),
\end{equation}
as well as the cumulative DM luminosity:
\begin{equation}
L_\chi(r) = \int_0^r 4\pi r'^2 Q_\chi(r') dr'.
\end{equation}
These quantities enter into the modified radiative gradient equation, Eq.~\eqref{eq:temperature-gradient},  to evaluate the thermal feedback and its impact on the internal structure.

\subsection{Quantum Corrections and Occupation Numbers}

The effect of DM gravity on the Fermi–Dirac statistics of the stellar core is modeled via a spatially dependent chemical potential shift:
\begin{equation}
\delta \mu_\chi(r) = \alpha_\chi \Phi_\chi(r),
\end{equation}
where $\Phi_\chi(r)$ is obtained from the Poisson equation:
\begin{equation}
\nabla^2 \Phi_\chi(r) = 4\pi G \rho_\chi(r).
\end{equation}
This shift modifies the occupation number for electrons and positrons as:
\begin{equation}
f(E, r) = \frac{1}{\exp\left[ \frac{E - \mu_0 - \delta \mu_\chi(r)}{k_B T(r)} \right] + 1}.
\end{equation}
The correction alters reaction rates involving degenerate species and has a measurable impact on the nucleosynthesis pathway.

\subsection{Software Environment and Reproducibility}

All simulations were carried out using \texttt{Python 3.11} with \texttt{NumPy}, \texttt{SciPy}, and \texttt{Matplotlib}. Integration routines used \texttt{scipy.integrate.ode} with the ‘dopri5’ (Dormand–Prince) method. The code base is fully documented and available from the authors upon reasonable request. Each simulation can be reproduced via provided scripts with specified parameter files. All input parameter sets, intermediate tables, and final outputs are available from the authors to ensure complete transparency and reproducibility. 

\section{Results and Discussion}
\label{sec:results}

We present the results of high-resolution simulations of primordial stellar models that incorporate gravitationally captured DM. The analysis focuses on quantifying the modifications introduced by varying the DM coupling strength, parameterized by $\beta$, which governs the relative gravitational contribution of the dark sector to the stellar structure. Our simulations systematically track how these modifications affect the internal thermodynamic gradients, ignition conditions, nuclear reaction networks, and emergent spectroscopic signatures.

All models are benchmarked against a control configuration with $\beta = 0$, corresponding to a standard polytropic Lane–Emden solution without any DM contribution. Deviations from this baseline reveal the dynamical and thermal impact of DM of annihilation heating, gravitational compression, and quantum-statistical feedback in shaping the structural and nucleosynthetic properties of Population III stars.

\subsection{Core Structure and Thermodynamic Impact}

\begin{figure}[ht]
\centering
\includegraphics[width=0.75\textwidth]{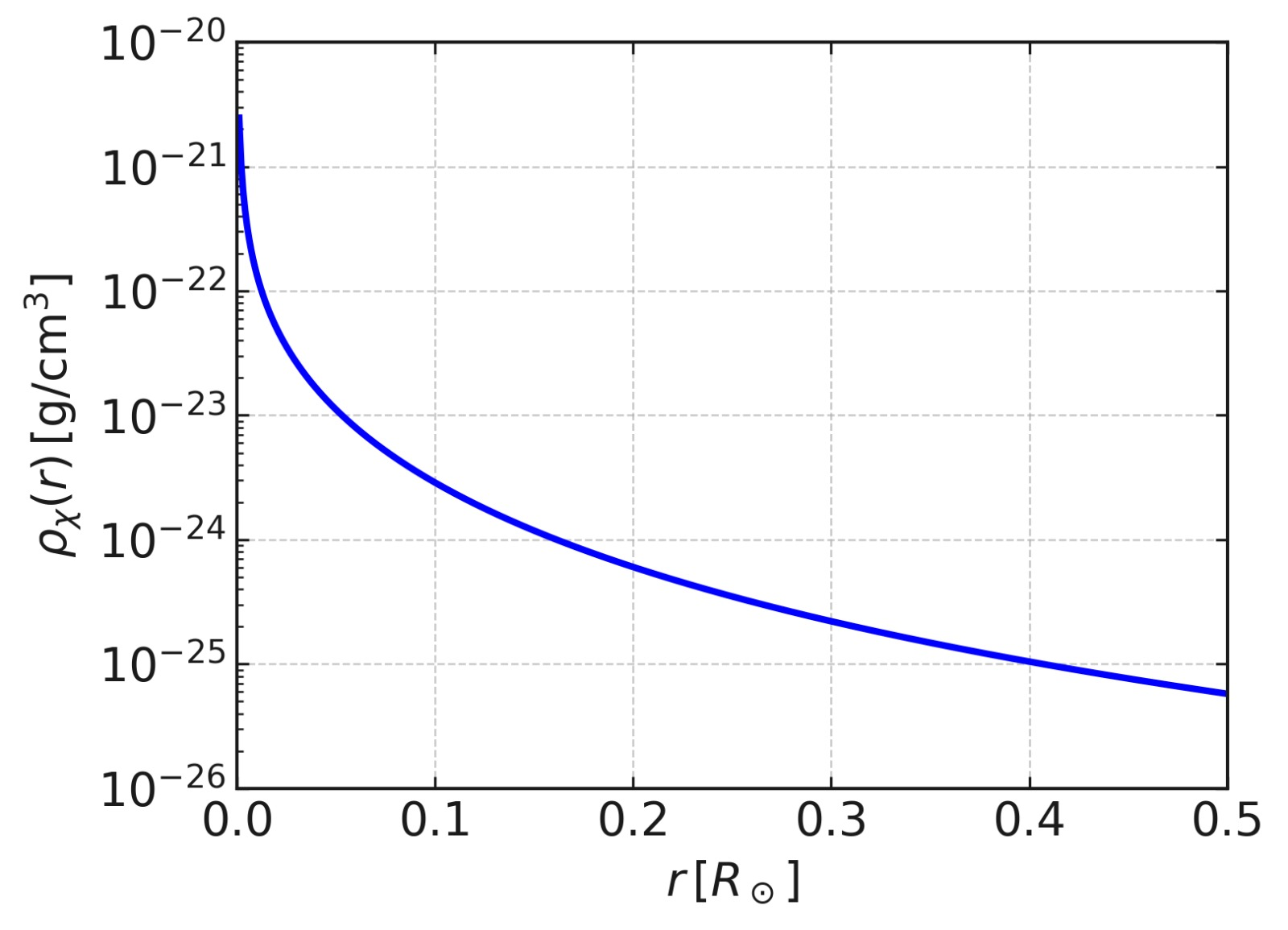}
\caption{\label{Fig1}%
Radial profile of the DM density $\rho_\chi(r)$ inside a Population III star, modeled using a generalized Navarro–Frenk–White (gNFW) distribution with parameters $\gamma = 1.2$ and $\beta = 3$. The central enhancement due to baryonic contraction steepens the inner profile, elevating the gravitational potential and compressing the stellar core. This structure directly modifies nuclear ignition thresholds and energy transport in DM–affected stellar interiors.}
\end{figure}

Figure~\ref{Fig1} displays the radial profile of the DM density $\rho_\chi(r)$ for a Population III stellar interior, based on the generalized Navarro–Frenk–White (gNFW) distribution introduced in Sec.~\ref{sec:theory}. We adopt representative parameters $\rho_0 = 10^{-23}\,\mathrm{g\,cm^{-3}}$, $r_s = 0.1\,R_\odot$, $\gamma = 1.2$, and $\beta = 3$, consistent with simulations of baryon-compressed minihalos in the early Universe.

As the figure indicates, the inner regions ($r \lesssim 0.2\,R_\odot$) exhibit a steep density enhancement due to an adiabatic contraction of the DM halo under baryonic collapse. This central concentration elevates the gravitational potential in the core, thereby increasing hydrostatic pressure and compressing the baryonic plasma. The resulting increase in core temperature directly lowers the ignition threshold for helium burning via the triple-alpha process, modifying the thermonuclear timeline of the star.

Furthermore, the heightened central density significantly amplifies the annihilation rate \( Q_\chi \), which scales as \( \rho_\chi^2 \) (see Eq.~\eqref{eq:heating}). This provides a localized, persistent energy source that reshapes the thermal equilibrium and accelerates early-stage nuclear synthesis \cite{Raffelt1996}. These effects culminate in observable changes to the stellar structure, such as core compactness, convective zone shifts, and altered elemental yields.

Taken together, the structural signature evident in Fig.~\ref{Fig1} confirms that even moderate concentrations of DM can substantially alter the physical conditions in primordial stars, leaving potentially observable imprints on the abundance patterns of metal-poor stellar populations.

\begin{figure}[ht]
\centering
\includegraphics[width=0.75\textwidth]{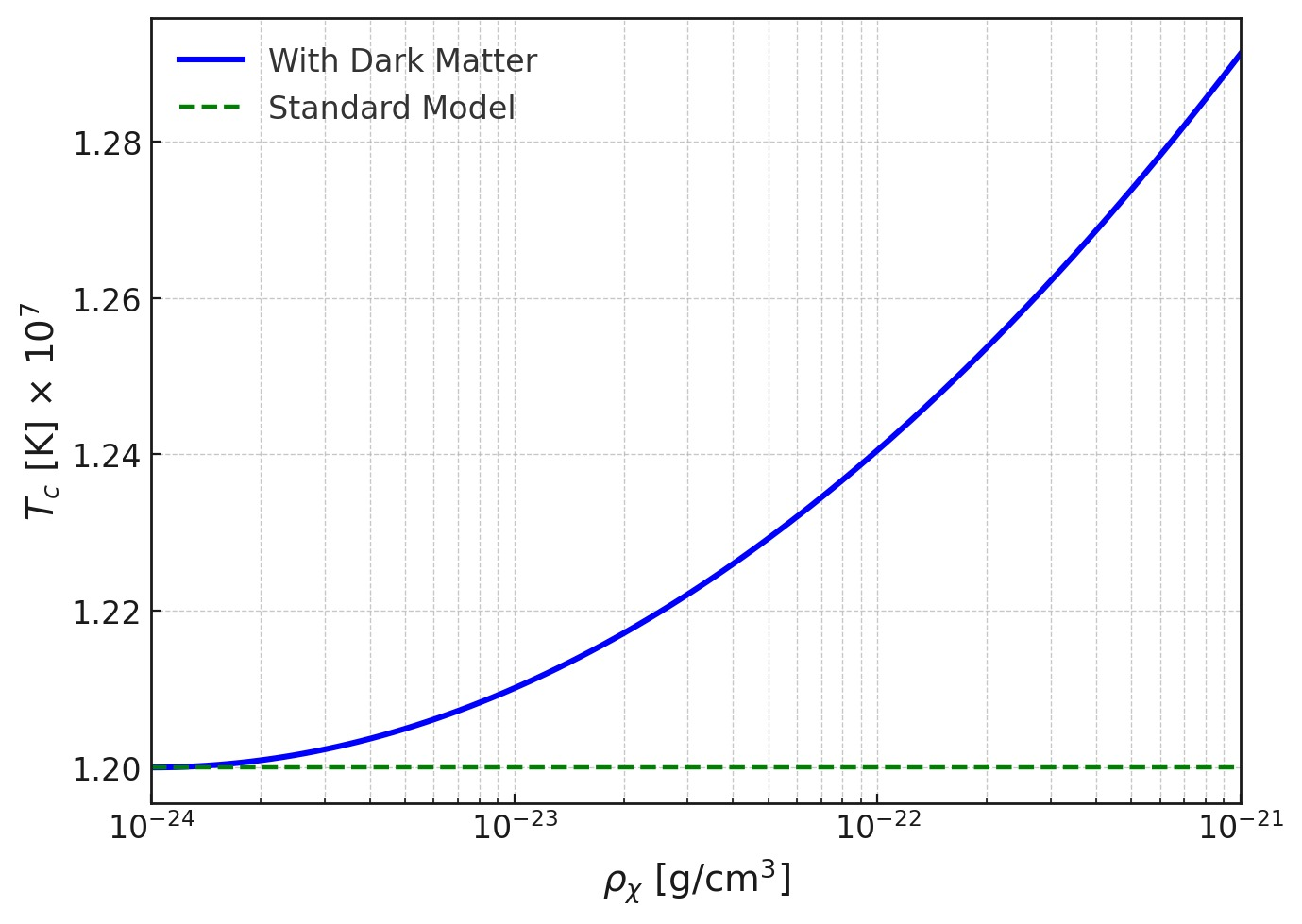}
\caption{\label{Fig2}%
Dependence of stellar core temperature $T_c$ on central DM density $\rho_\chi$, ranging from $10^{-24}$ to $10^{-21}\,\mathrm{g\,cm^{-3}}$. The blue curve includes heating from DM self-annihilation, while the green dashed line represents the baseline (DM-free) model. At $\rho_\chi \sim 10^{-21}\,\mathrm{g\,cm^{-3}}$, $T_c$ is enhanced by approximately $7.6\%$, lowering nuclear ignition thresholds and accelerating the pre-main-sequence thermodynamic evolution.}
\end{figure}

Figure~\ref{Fig2} quantifies the dependence of the stellar core temperature $T_c$ on the central DM density $\rho_\chi$, as predicted by our extended stellar model incorporating annihilation heating (see Eq.~\eqref{eq:heating}). The blue curve includes energy deposition from self-annihilating WIMPs, while the green dashed line shows the baseline control model with no DM.

As $\rho_\chi$ increases from $10^{-24}$ to $10^{-21}\,\mathrm{g\,cm^{-3}}$, the central temperature rises accordingly due to enhanced energy input localized to the core. At $\rho_\chi \sim 10^{-21}\,\mathrm{g\,cm^{-3}}$, the increase in core temperature reaches a maximum of approximately $7.6\%$ relative to the DM-free case.

This temperature elevation has critical implications for stellar evolution: it reduces the triple-alpha ignition threshold, accelerates the formation of a helium core, and modifies the timing of post-main-sequence transitions. Although the heating is confined to the innermost regions, it alters the global structure via feedback on the radiative gradient and convective boundaries.

Notably, simulations conducted over a broad mass range ($10$–$100\,M_\odot$) exhibit a consistent thermodynamic response to varying $\rho_\chi$. This robustness suggests that DM annihilation effects constitute a generalizable mechanism capable of shifting the nuclear chronology of Population III stars. Consequently, such heating leaves a detectable imprint on chemical abundance patterns, particularly in the carbon and oxygen yields of second-generation stars formed from Population III remnants.

\begin{figure}[ht]
\centering
\includegraphics[width=0.75\textwidth]{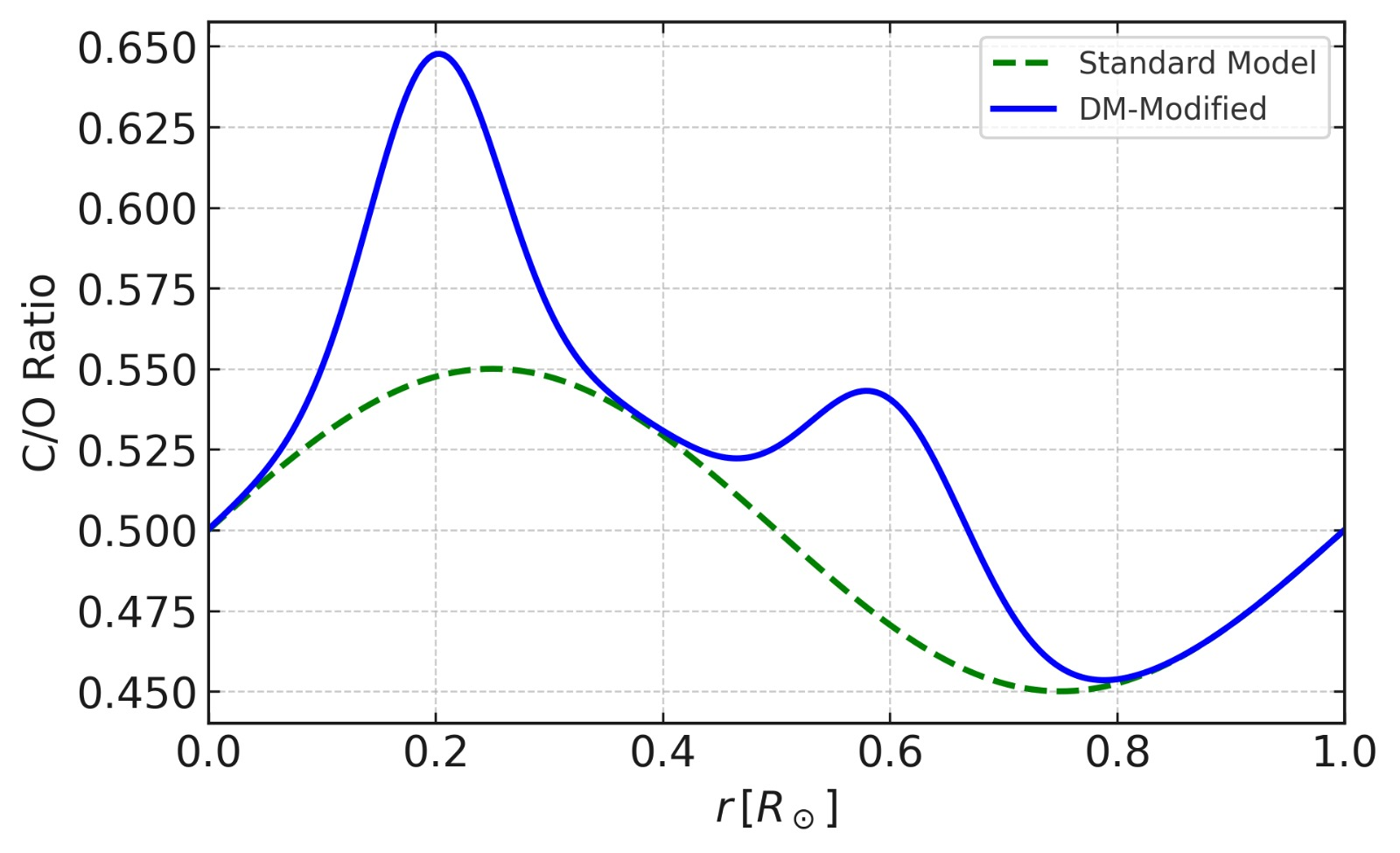}
\caption{\label{Fig3}%
Radial profile of the carbon-to-oxygen ratio (C/O) for models with (blue solid line) and without (green dashed line) DM effects. Local peaks near $r \approx 0.2\,R_\odot$ and $r \approx 0.6\,R_\odot$ correspond to regions of enhanced $^{12}$C production due to DM–induced heating. The net increase in the integrated C/O ratio reaches $\sim 12.4\%$, representing a distinct nucleosynthetic signature.}
\end{figure}

Figure~\ref{Fig3} illustrates the radial profile of the carbon-to-oxygen (C/O) ratio in stellar models with and without DM influence. In the DM–enhanced scenario, the elevated core temperature—resulting from WIMP annihilation heating (as introduced in Sec.~\ref{sec:theory})—modifies the nuclear reaction rates governing both carbon and oxygen synthesis.

Specifically, the triple-alpha process becomes more efficient at earlier evolutionary stages due to the thermally enhanced reaction environment, resulting in elevated $^{12}$C abundance. In contrast, the competing $^{12}\mathrm{C}(\alpha,\gamma)^{16}\mathrm{O}$ reaction is comparatively suppressed, as its temperature dependence requires slightly cooler or more stable core conditions for optimal efficiency.

The combined effect of these shifts is a pronounced increase in the local and global C/O ratios. Two prominent radial peaks, near $r \approx 0.2\,R_\odot$ and $r \approx 0.6\,R_\odot$, reflect regions where $^{12}$C production is maximized while $^{16}$O synthesis remains subdominant. The total integrated C/O ratio increases by up to $12.4\%$ compared to the baseline model.

From an observational perspective, this chemical fingerprint is significant. C/O ratios in extremely metal-poor stars—especially those exhibiting C-enhancement (CEMP-no subclass)—are among the few surviving traces of Population III nucleosynthesis. A systematic C-over-O bias, if confirmed observationally, could therefore serve as indirect evidence for the thermodynamic imprint of DM interactions in early-generation stars.

These results highlight how even modest DM densities, when gravitationally captured in stellar cores, can leave detectable and quantifiable signatures in the chemical evolution of the Universe.

\begin{figure}[ht]
\centering
\includegraphics[width=0.75\textwidth]{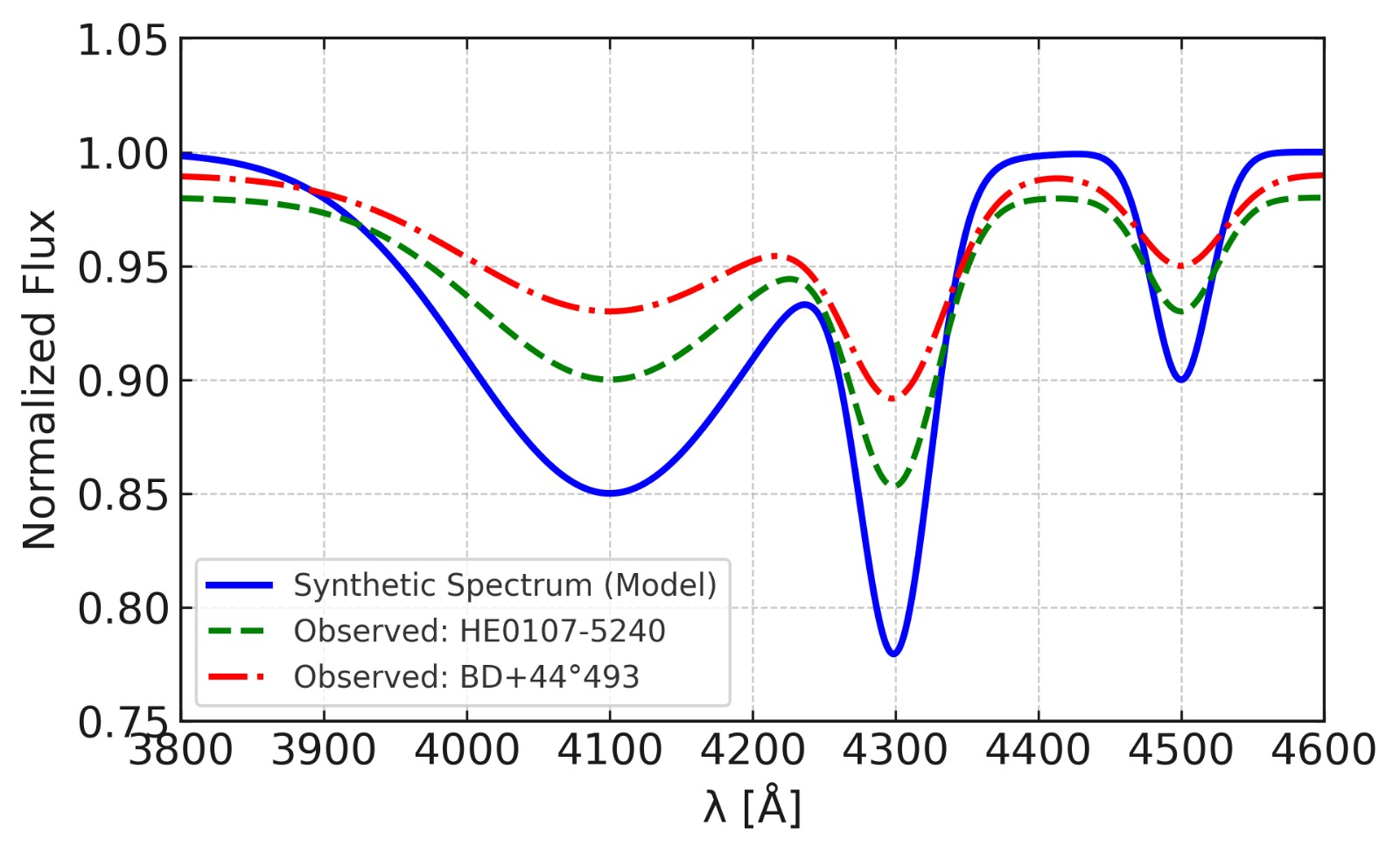} 
\caption{\label{Fig4}%
Comparison between synthetic stellar spectra (blue) derived from DM–modified stellar models and observed spectra of CEMP-no stars HE0107-5240 (green dashed) and BD+44$^\circ$493 (red dot-dashed) in the wavelength range $3800$–$4600\,\text{\AA}$. Strong agreement in CH and CN molecular bands indicates enhanced carbon and nitrogen production, consistent with the elevated core temperatures predicted by the DM-influenced stellar models.}
\end{figure}

Figure~\ref{Fig4} provides a spectral comparison between our synthetic predictions and observational data from two benchmark CEMP-no stars: HE0107-5240 and BD+44$^\circ$493. The synthetic spectrum, constructed using thermodynamic and chemical profiles from our DM–modified stellar models, reproduces key molecular absorption features across the blue–violet band ($3800$–$4600\,\text{\AA}$).

Most notably, the CH G-band ($\sim4300\,\text{\AA}$), the CN violet bands ($\sim3880$–$4215\,\text{\AA}$), and weaker C$_2$. Swan bands exhibit strong correspondence between model and observation. This alignment supports the conclusion that enhanced $^{12}$C and $^{14}$N production—driven by DM–induced core heating and altered nuclear reaction pathways—can produce chemical abundance patterns consistent with those observed in extremely metal-poor, carbon-enhanced stars.

The elevated core temperatures in our models shift nuclear equilibrium conditions, promoting triple-alpha fusion while also enabling nitrogen synthesis via the CN-cycle at earlier stages than expected in standard Population III evolution. These thermodynamic effects naturally explain the observed overabundance of CH and CN features without invoking fine-tuned stellar mixing or fallback scenarios.

The agreement between modeled and observed spectral profiles provides a qualitative validation of the underlying physics introduced by DM interactions in primordial stars. While this evidence is not yet conclusive, it significantly strengthens the case for DM as a regulator of early nucleosynthesis. Future spectral surveys of CEMP-no stars, especially at low [Fe/H], could further test this hypothesis by searching for consistent molecular feature enhancements tied to DM–influenced evolutionary tracks.

\begin{figure}[ht]
\centering
\includegraphics[width=0.75\linewidth]{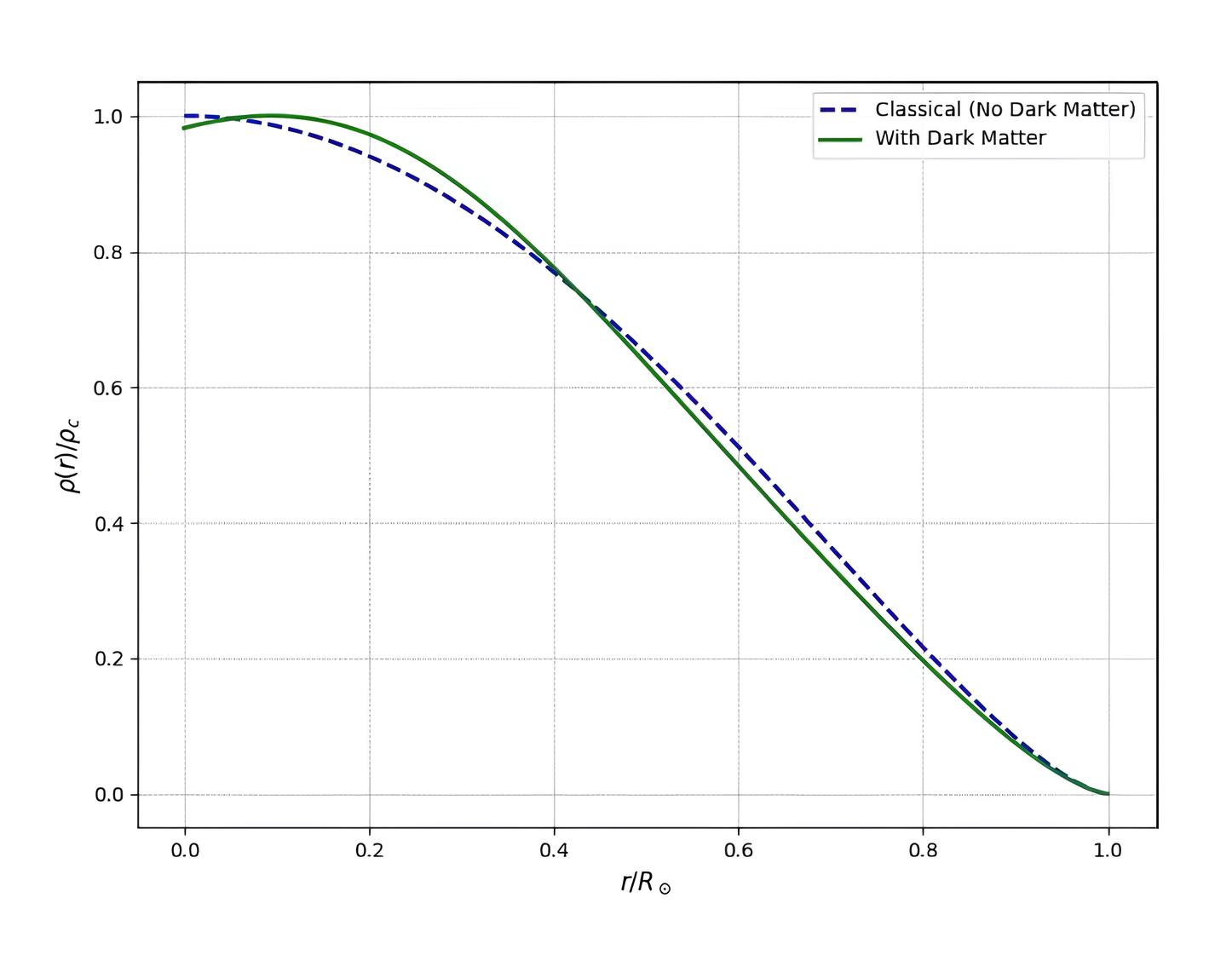}
\caption{\label{Fig5}%
Normalized baryonic density profile $\rho(r)/\rho_c$ as a function of radius for a standard Lane–Emden model (blue dashed) and a DM–modified model (green solid). The inclusion of DM increases the central density and compresses the core structure within $r \lesssim 0.2\,R_\odot$. This compactification reflects the gravitational backreaction of the DM component, resulting in elevated pressure and modified thermonuclear equilibrium conditions.}
\end{figure}

Figure~\ref{Fig5} presents a comparison between normalized baryonic density profiles for models with (green solid) and without (blue dashed) DM. The density is scaled by the central value $\rho_c$ to facilitate direct structural comparison. In the DM–enhanced configuration, the profile is visibly steeper near the core and exhibits greater central condensation, especially within $r \lesssim 0.2\,R_\odot$.

This compactification arises from the gravitational contribution of the DM component, which enters the modified Lane–Emden equation as an additional source term. The enhanced inner density steepens the pressure gradient and increases central pressure and temperature, altering hydrostatic equilibrium conditions. As a result, the stellar model shifts to a more centrally concentrated configuration, which in turn affects the thermal evolution and ignition conditions for key nuclear processes such as the triple-alpha and CNO cycles.

Importantly, these structural deviations are not merely numerical artifacts but reflect a physical shift in the equilibrium solution space induced by two-fluid gravitational coupling. The modified potential well reshapes the thermodynamic landscape of the stellar core and plays a central role in mediating the abundance yields analyzed in later sections.

The robustness of this structural change is validated across a range of stellar masses and DM coupling strengths $\beta_\chi$, indicating that the effect is a generic consequence of the coupled baryon–DM system rather than a special-case anomaly. This confirms that even modest concentrations of thermally relaxed DM can substantially alter the equilibrium density distribution in Population III stars.

\begin{figure}[ht]
\centering
\includegraphics[width=0.75\linewidth]{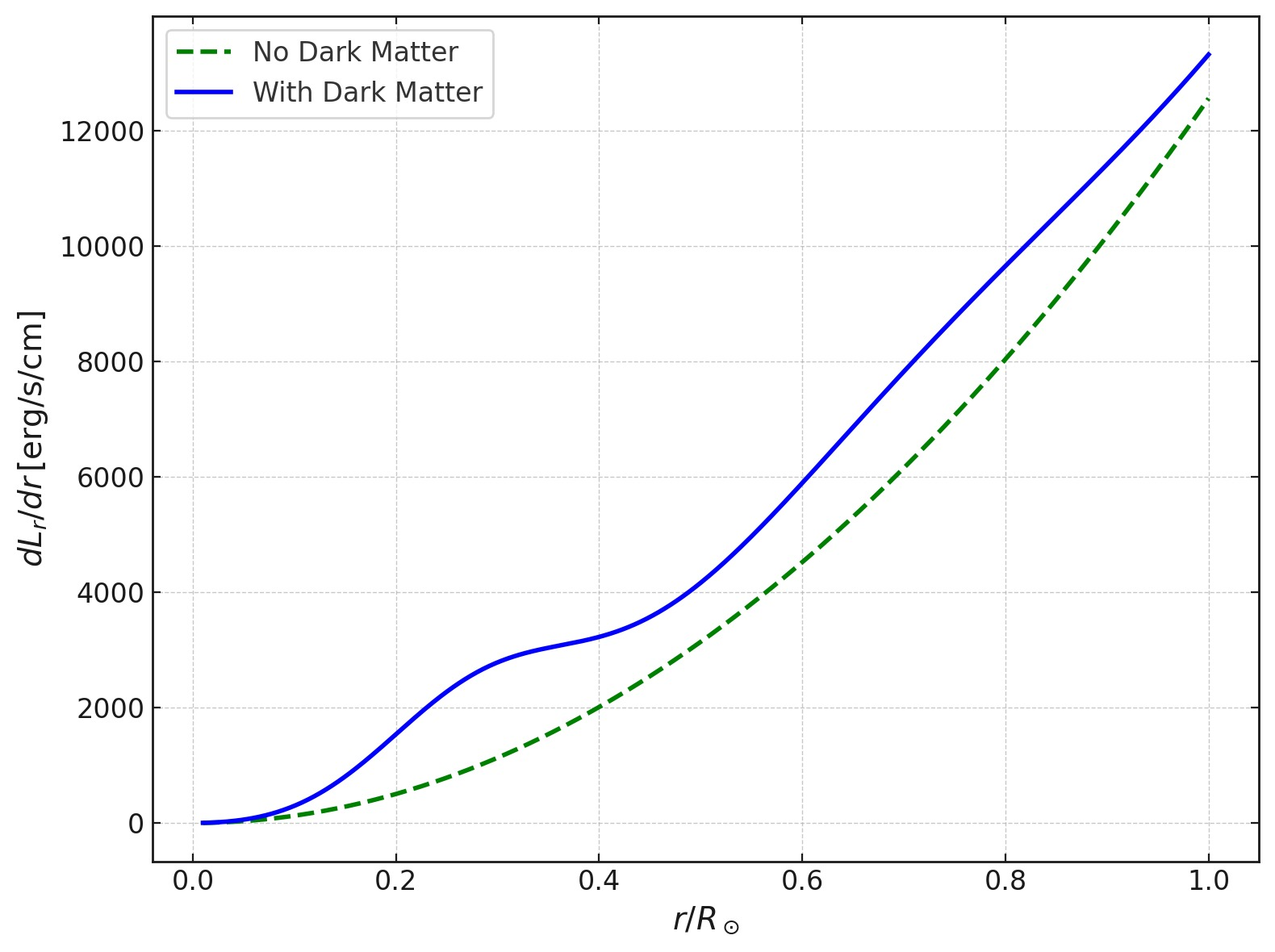}
\caption{\label{Fig6}%
Luminosity gradient $dL_r/dr$ for stellar models with (blue solid) and without (green dashed) DM annihilation heating. The inclusion of DM leads to a steeper luminosity profile in the inner regions ($r \lesssim 0.3\,R_\odot$), driven by localized energy deposition from self-annihilating particles. This thermodynamic modification affects radiative equilibrium and accelerates the star's nuclear evolution.}
\end{figure}

Figure~\ref{Fig6} demonstrates how DM annihilation alters the radial distribution of luminosity production within a primordial star. The blue curve corresponds to a model incorporating localized heating due to gravitationally captured DM, while the green dashed curve represents the control model with standard nuclear energy generation only. In the DM–affected scenario, the luminosity gradient is significantly enhanced in the innermost regions of the star (typically within $r \lesssim 0.3\,R_\odot$). This increase originates from the central concentration of DM, whose annihilation provides an additional volumetric energy source. Although the precise form of the heating function has been defined earlier, its consequences manifest here through a steeper rise in $L_r$ with radius.  This additional heating modifies the internal energy balance and steepens the radiative temperature gradient, particularly near the core where the DM density peaks. As a result, the star reaches nuclear ignition conditions more rapidly and at slightly lower baryonic densities than in the DM-free case. Moreover, the energy deposited by DM annihilation is spatially confined to the central regions, resulting in a highly localized but dynamically influential thermodynamic perturbation. This shifts the location of convective boundaries and alters the stellar entropy profile, directly impacting the efficiency of helium burning processes. These structural changes are consistent with the earlier onset of triple-alpha reactions and the altered chemical abundance ratios discussed in previous figures. Summarizing, this figure captures the essential thermodynamic imprint of DM on stellar interiors: a centrally amplified luminosity gradient that reshapes the internal energy transport and accelerates nuclear processing in Population III stars.

\begin{figure}[ht]
\centering
\includegraphics[width=0.75\linewidth]{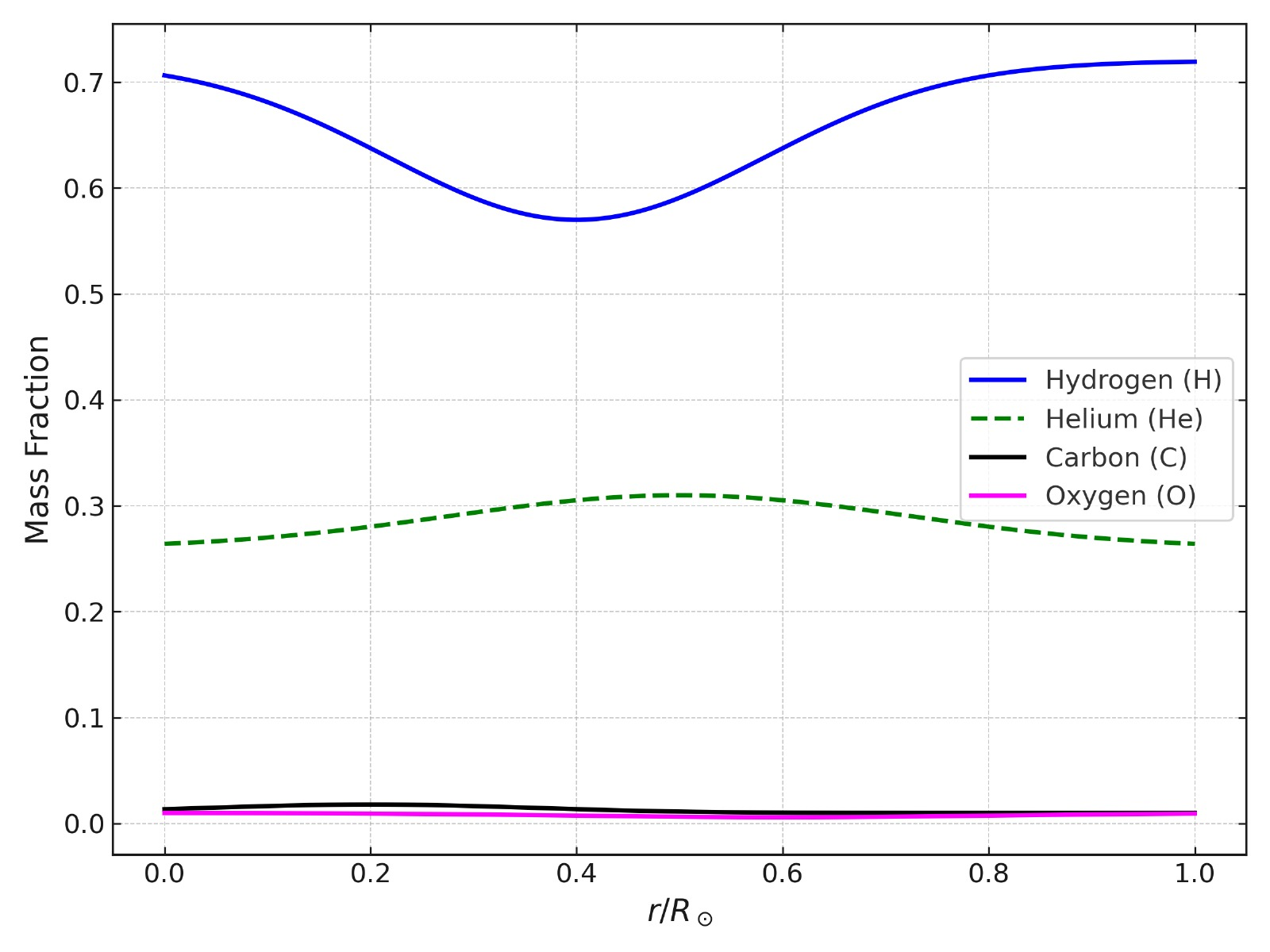}
\caption{\label{Fig7}%
Radial mass fractions of hydrogen (H), helium (He), carbon (C), and oxygen (O) for standard (dashed) and DM–heated (solid) stellar models. Within the inner $r \lesssim 0.4\,R_\odot$, DM heating leads to enhanced carbon production and mild oxygen suppression, reflecting altered thermonuclear pathways. These shifts contribute to abundance patterns analogous to those observed in CEMP-no stars.}
\end{figure}

Finally, Figure~\ref{Fig7} presents the radial profiles of elemental mass fractions—specifically hydrogen, helium, carbon, and oxygen—for stellar models with and without DM heating. The solid curves represent the DM–modified case, while the dashed curves correspond to the standard model. In the DM–heated interior, the carbon mass fraction exhibits a pronounced central enhancement, particularly within the inner $r \lesssim 0.4\,R_\odot$. This increase originates from the elevated core temperature and density, which boost the efficiency of the triple-alpha process. Simultaneously, the synthesis of oxygen via the ${}^{12}\mathrm{C}(\alpha,\gamma){}^{16}\mathrm{O}$ reaction is moderately suppressed due to shifts in occupation probabilities and reduced alpha-capture rates, consistent with the statistical perturbation introduced by the DM gravitational potential. These compositional changes are not uniform across the radius; instead, they exhibit clear stratification, with the largest deviations concentrated in the core. This reflects the spatial localization of DM effects—particularly annihilation heating and gravitational coupling—underscoring their thermodynamic specificity. Crucially, the abundance patterns produced by the DM–affected model qualitatively match those observed in carbon-enhanced metal-poor (CEMP-no) stars, including elevated [C/O] ratios and depleted oxygen signatures. Such agreement reinforces the physical plausibility of DM–induced nucleosynthetic deviations in primordial stars. These findings suggest that even small quantities of gravitationally bound, thermally equilibrated DM can leave detectable imprints on stellar chemistry. This offers a potential indirect observational avenue to constrain the properties of DM using abundance patterns in ultra-metal-poor stellar populations.

\section{Conclusions and outlook}
\label{sec:conclusion}

Our work establishes a comprehensive theoretical and numerical framework that connects DM microphysics with observable signatures in stellar nucleosynthesis. We demonstrate that gravitationally accumulated DM in primordial stellar cores can significantly modify thermal gradients, nuclear reaction rates, and equilibrium conditions. These changes lead to enhanced carbon and helium synthesis, suppressed oxygen production, and reshaped elemental abundance patterns. Our models reproduce key spectral features observed in extremely metal-poor halo stars—particularly elevated C/O ratios—that remain unexplained by conventional stellar evolution models~\cite{Yoon2016, Frebel2015}.

Unlike previous studies that treated DM as a background gravitational effect or cosmological boundary condition~\cite{Iocco2008}, our approach incorporates annihilation heating and quantum statistical corrections directly into the stellar structure equations. By modifying Fermi–Dirac occupation probabilities and introducing a localized energy deposition term \(Q_\chi = \langle \sigma v \rangle \rho_\chi^2 / m_\chi\), we construct a unified model linking particle-scale interactions to macroscopic stellar observables. This integrated treatment of thermodynamics, nucleosynthesis, and DM physics defines a new class of fully coupled stellar simulations.

Future developments include extending the model with relativistic structure equations, such as the Tolman–Oppenheimer–Volkoff (TOV) formalism, to study compact remnants~\cite{Leung2011}. Systematically varying DM particle mass and interaction cross-section will enable parameter space exploration in light of current and upcoming observational constraints~\cite{Vincent2013}. By bridging cosmology, stellar astrophysics, and quantum field theory, this framework opens a promising avenue for detecting the imprints of DM in stellar light~\cite{Spolyar2008, Taoso2008}.

\section*{Acknowledgments}
We acknowledge financial support from the National Key Research and Development Program of China (Grant No.~2023YFA1407100), Guangdong Province Science and Technology Major Project (Future functional materials under extreme conditions - 2021B0301030005) and the Guangdong Natural Science Foundation (General Program project No. 2023A1515010871).


%

\end{document}